\documentclass[aps,prb,twocolumn,groupedaddress,showpacs]{revtex4}

\usepackage{graphicx}

\begin{document}

\title{Bell test with time-delayed two-particle correlations}

\author{A.V.\ Lebedev$^{\, a,b}$ and G.\ Blatter$^{\, a}$}

\affiliation{$^{a}$Theoretische Physik, Schafmattstrasse 32,
ETH-Zurich, CH-8093 Z\"urich, Switzerland}

\affiliation{$^{b}$L.D.\ Landau Institute for Theoretical Physics,
RAS, 119334 Moscow, Russia}

\date{\today}

\begin{abstract}
   Adopting the frame of mesoscopic physics, we describe a Bell type
   experiment involving time-delayed two-particle correlation measurements.
   The indistinguishability of quantum particles results in a specific
   interference between different trajectories; the non-locality in the
   time-delayed correlations manifests itself in the violation of a Bell
   inequality, with the degree of violation related to the accuracy of the
   measurement.  In addition, we demonstrate how the interrelation between the
   orbital- and the spin-exchange symmetry can by exploited to infer knowledge
   on spin entanglement from a measurement of orbital entanglement.
\end{abstract}


\maketitle

\section{Introduction}

\begin{figure}[bt]
   \includegraphics[width=7.5cm]{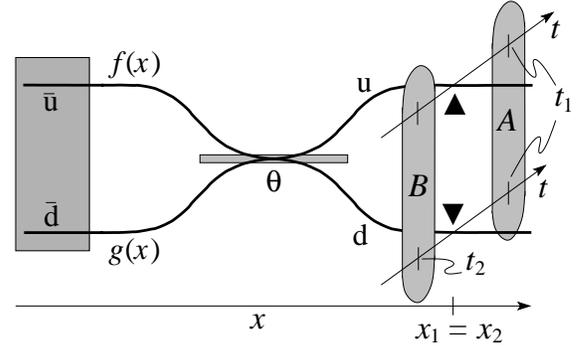} \caption[]{Particles incident in
   leads $\bar\mathrm{u}$ and $\bar\mathrm{d}$ with wave functions $f(x)$ and
   $g(x)$ are mixed in a four-terminal splitter (characterized by the mixing
   angle $\theta$) and analyzed at $x_1 = x_2$ through measurement of
   time-correlations during the time intervals $t_1 \in A$ and $t_2 \in B$
   (time axis drawn perspectively into the plane). We are interested in the
   entanglement of the lead indices u and d with respect to bipartitioning of
   the system between the time intervals $A$ and $B$.  For details on the
   implementation of the source (shaded area) see Fig.\ \ref{fig:source}.}
   \label{fig:setup_a}
\end{figure}

Fundamental quantum phenomena, such as non-locality and entanglement of
quantum degrees of freedom, have regained a lot of interest recently, mainly
due to their potential usefulness as a computational resource. Mesoscopic
physics provides a new platform for the investigation of these phenomena,
important issues being the creation, quantification, and verification of
non-locality/entanglement. In this paper, we describe an experiment where two
electrons with different orbital wave functions are superposed in an
interferometer and analyzed in a Bell type experiment involving two-particle
correlation measurements, see Fig.\ \ref{fig:setup_a}. The particular feature
of this Bell test is the replacement of the four different settings of local
detectors in the original setup by four different time-delays in the measured
correlators. The main physical property we want to exploit is the
indistinguishability of quantum particles, which results in a specific
interference between different trajectories. We wish to convey three
messages: first, the non-locality in the time-delayed correlations due to
indistinguishability manifests itself in the violation of a Bell inequality.
Second, the degree of violation is related to the accuracy of the measurement
and is reduced, once the local measurement can distinguish between the different
orbital wave functions of the particles.  The above two items refer to
spinless objects (or particles with equal spin).  Third, adding the spin
degree of freedom, we show how the symmetry relation between spin- and orbital
components allows to extract information on spin-entanglement from an orbital
measurement.

By now, numerous proposals have been made how to create entangled states in
mesoscopic setups, both for orbital- and spin degrees of freedom
\cite{review}.  The verification and quantification of entanglement can be
carried out using Bell inequality checks \cite{cht_00} or state tomography
\cite{loss_03,buttiker,fazio}.  The indistinguishability of (spinless)
particles producing a two-particle Aharonov-Bohm effect and entanglement has
been exploited in a Hanbury-Brown Twiss interferometer
\cite{SamBut_03,Heiblum_07}; here, we offer an alternative implementation
which makes use of an electron beam splitter. Using the same setup as
discussed here, Burkhard {\it et al.}\cite{loss_00} have demonstrated how to
distinguish between singlet and triplet spin-states by a measurement of
zero-frequency cross-correlations. Below, we exploit that the orbital
measurement of the Bell parameter preserves the spin-entanglement; this
feature allows us to find a lower bound on the concurrence of the spin wave
function.

In the following, we consider a setup with two incoming leads, denoted as
$\bar \mathrm{u}$ and $\bar \mathrm{d}$, connected to two outgoing leads
$\mathrm{u}$ and $\mathrm{d}$ through a reflectionless four-terminal beam
splitter, see Fig.\ \ref{fig:setup_a}. At time $t=0$ two electrons with normalized
orbital wave functions $f(x)$ and $g(x)$ and common spin state
$\chi(\sigma_1,\sigma_2)$ are injected into the leads $\bar\mathrm{u}$ and
$\bar \mathrm{d}$. The state, factorizable in orbital and spin parts (with
factorized orbital part and general spin part), is conveniently written within
a second quantized formalism,
\begin{eqnarray}
      &&|\Psi_\mathrm{in}\rangle =
      \int dx_1 dx_2\, f(x_1) g(x_2)
      \label{inst}
      \\
      &&\qquad \times
      \sum_{\sigma_1\sigma_2} \chi(\sigma_1,\sigma_2)\,
      \hat \psi_{\bar\mathrm{u}\sigma_1}^\dagger(x_1)
      \hat \psi_{\bar\mathrm{d}\sigma_2}^\dagger(x_2) | 0\rangle;
      \nonumber
\end{eqnarray}
here,  $\hat \psi_{\alpha\sigma}^\dagger(x)$ creates electrons at
the position $x$ in lead $\alpha$ and $|0\rangle$ is the vacuum
state with no electrons. After mixing in a four-terminal splitter
(with mixing angle $\theta$), we will analyze correlations in the
system through detection of particles in time (see Fig.\
\ref{fig:setup_a}) or space separated intervals $A$ and $B$ (see
Fig.\ \ref{fig:setup_b}). Our focus then is on the entanglement of
the lead indices u and d with respect to bipartitioning of the
system between the time or space intervals $A$ and $B$.

\section{Bell test}

The particles in the outgoing leads $\mathrm{u}$ and $\mathrm{d}$ are
subjected to a Bell test expressed through time-resolved current-current
correlators in the leads $\alpha_1$ and $\alpha_2$, $\alpha_1,\alpha_2 \in
\{\mathrm{u},\mathrm{d}\}$ (both auto- $\alpha_1 = \alpha_2$ and crossed-
$\alpha_1 \neq \alpha_2$ correlators are considered),
\begin{equation}
      C_{\alpha_1\alpha_2}(AB) = \frac1{\delta t^2} \int_A dt_1
      \int_B dt_2 \, \langle \hat I_{\alpha_1}(x_1,t_1) \hat
      I_{\alpha_2}(x_2,t_2)\rangle,
      \label{corr}
\end{equation}
where $\hat I_\alpha(x,t)$ is the total current operator (summed over spin
degrees of freedom) in lead $\alpha$ at position $x$ and time $t$.  The time
integration is taken over a finite time interval $A =[t_A-\delta t/2, t_A
+\delta t/2]$ (same for $B$) with the width $\delta t$ accounting for the
finite time-resolution of the current measurement, cf.\ Fig.\
\ref{fig:setup_a}; the limit $\delta t\rightarrow 0$ corresponds to a
measurement of the instantaneous current. In the following, we will assume
that all correlators are measured at some fixed symmetric position $x_1=x_2$
and omit the coordinate variable.

With only two electrons present in the system and for non-overlapping
time-intervals $A\cap B = 0$, the correlation function $C_{\alpha_1
\alpha_2}(AB)$ is proportional to the joint probability $P_{\alpha_1
\alpha_2}(AB)$ for the detection of two particles during the time intervals
$A$ and $B$ in the leads $\alpha_1$ and $\alpha_2$, see Ref.\
\onlinecite{Lebedev_05}.  There are four distinct possibilities to distribute
two electrons between the outgoing leads and we can define the properly
normalized ($\sum_{\alpha_1\alpha_2} P_{\alpha_1\alpha_2}(AB)=1$)
probabilities as
\begin{equation}
   P_{\alpha_1\alpha_2}(AB) = \frac{C_{\alpha_1\alpha_2}(AB)}
   {\sum_{\alpha_1\alpha_2} C_{\alpha_1\alpha_2}(AB)}.
\label{prob}
\end{equation}
Out of these, we define the two-particle Bell inequality in the
Clauser-Horne~\cite{clauser} form in the same way as it is done in the usual
optics context \cite{aspect}: we introduce the Bell correlation functions
\begin{equation}
      E_{AB} = [P_\mathrm{uu} - P_\mathrm{ud} -
      P_\mathrm{du} + P_\mathrm{dd}]_{AB}
      \label{Bcorr}
\end{equation}
and obtain the Bell inequality
\begin{equation}
      \bigl| E_{AB} - E_{AB'} + E_{A' B} + E_{A' B'}
      \bigr| \leq 2.
      \label{BI}
\end{equation}
Here, the polarizations $\pm$ in the optics context are replaced
by the lead indices $\mathrm{u}$ and $\mathrm{d}$ and the role of
the four different polarization settings of the detectors is played
by four different time intervals $A$, $B$ and $A'$, $B'$.
The violation of this inequality for a particular choice of time
intervals shows that non-local correlations are present in the
system, i.e., the result of the measurement cannot be simulated by
any local-variable theory.

Let us demonstrate that the above Bell inequality indeed can be violated by
the incoming state~(\ref{inst}) after proper projection. We then have to
calculate the four current-current auto- and cross-correlators
$C_{\alpha_1\alpha_2}$ with $\alpha_1= \alpha_2$ and $\alpha_1\neq \alpha_2$,
respectively.  This is done within the scattering matrix approach to quantum
noise~\cite{transport}: we assume that the Fourier components $f(k)$ and
$g(k)$ of the single-particle wave functions are concentrated near the wave
vector $k_0>0$, allowing us to linearize the energy--momentum dispersion near
$k_0$. The time evolution of the incoming state~(\ref{inst}) then is described
by the propagation of the single-particle wave packets $f(x)$ and $g(x)$ with
constant velocity $v_0=\hbar k_0/m$ to the right, $f(x,t) = f(\xi)$ and
$g(x,t)=g(\xi)$, where $\xi=x-v_0 t$ is a retarded variable. With the
scattering matrix of the beam splitter (parametrized by the angle $\theta$),
\begin{equation}
      \left( \begin{array}{c} \mathrm{u}\\ \mathrm{d} \end{array}
      \right) = \left( \begin{array}{cc} \cos\theta& -\sin\theta
      \\ \sin\theta & \cos\theta \end{array} \right) \left(
      \begin{array}{c} \bar \mathrm{u} \\ \bar \mathrm{d}
      \end{array} \right),
      \label{scm}
\end{equation}
we can express the current operators $\hat I_\alpha(x,t)$ in the outgoing
leads $\alpha \in \{\mathrm{u},\mathrm{d}\}$ through the electronic scattering
states. Averaging the product of current operators in Eq.~(\ref{corr}) over
the incoming state $|\Psi_\mathrm{in}\rangle$ one arrives at the results
\begin{eqnarray}
      &&\langle \hat I_\mathrm{u}(\xi_1) \hat I_\mathrm{u} (\xi_2)
      \rangle = (ev_0)^2 \bigl\{\bigl(
      \cos^2\theta\,|f(\xi_1)|^2
      \label{auto}
      \\
      &&\qquad\qquad\qquad\qquad\qquad + \sin^2\theta\,|g(\xi_2)|^2 \bigr)
      \delta(\xi_1-\xi_2)
      \nonumber\\
      &&\quad + \sin^2\theta
      \cos^2\theta\,\bigl[|f(\xi_1)|^2|g(\xi_2)|^2+|g(\xi_1)|^2|f(\xi_2)|^2
      \nonumber\\
      &&\qquad\qquad\qquad
      -Q\,\bigl(f(\xi_1)g^*(\xi_1)g(\xi_2)f^*(\xi_2)+c.c.\bigr)\bigr]\bigr\},
      \nonumber \\
      && \langle \hat I_\mathrm{u}(\xi_1) \hat I_\mathrm{d} (\xi_2)
      \rangle = (ev_0)^2\bigl\{\cos^4\theta\, |f(\xi_1)|^2 |g(\xi_2)|^2
      \label{cross} \\
      && \qquad\qquad\qquad\qquad\qquad + \sin^4\theta\,|g(\xi_1)|^2 |f(\xi_2)|^2
      \nonumber\\
      &&\quad+ \sin^2\theta \cos^2\theta\,Q\,\bigl(f(\xi_1) g^*(\xi_1)
      g(\xi_2) f^*(\xi_2) + c.c. \bigr)\},
      \nonumber
\end{eqnarray}
where $Q = \sum_{\sigma_1\sigma_2} \chi(\sigma_1, \sigma_2) \chi^*(\sigma_2,
\sigma_1)$ describes the overlap between the spin states of the two electrons
in the incoming state~(\ref{inst}). The two other correlation functions
$\langle \hat I_\mathrm{d}(\xi_1) \hat I_\mathrm{d}(\xi_2) \rangle$ and
$\langle \hat I_\mathrm{d}(\xi_1) \hat I_\mathrm{u}(\xi_2) \rangle$ are
obtained by exchanging $\cos\theta$ and $\sin \theta$ in Eqs.~(\ref{auto})
and~(\ref{cross}).  Substituting these expressions for the current correlators
into Eq.~(\ref{corr}) and integrating over (non-overlapping) time intervals
$A$ and $B$, one arrives at the Bell correlation function Eq.~(\ref{Bcorr})
\begin{equation}
   E_{AB}=-\cos^2(2\theta)-\sin^2(2\theta)\,Q\,\frac{S_A S_B^* + S_A^* S_B}
   {F_A G_B + G_A F_B},
   \label{Bcorr2}
\end{equation}
where we have introduced the particle densities $F_{A,B}$ and $G_{A,B}$
averaged over the time intervals $A$ and $B$,
\begin{equation}
      F_{A,B} = \frac1{\delta t} \int_{t\in A,B} dt \, |f(\xi)|^2
\end{equation}
(and similarly for $G_{A,B}$ with $f$ replaced by $g$). The overlap $S_{A,B}$
between different single particle wave functions reads
\begin{equation}
      S_{A,B} = \frac1{\delta t} \int_{t\in A, B} dt \, f(\xi) g^*(\xi).
\end{equation}
In the following, we apply the result (\ref{Bcorr2}) first to
spinless fermions and plane wave states $f$ and $g$ and confirm
the violation of the Bell inequality in this simple situation. We
then proceed with a rederivation of the expression (\ref{Bcorr2})
with space-like separated measurement intervals $A$ and $B$ in
order to make the origin of the entanglement more transparent. A
formulation in terms of reduced density matrices leading to an
expression of the Bell correlator in terms of concurrences
completes the discussion.

\subsection{Spinless fermions}

\begin{figure}[bt]
   \includegraphics[width=7.5cm]{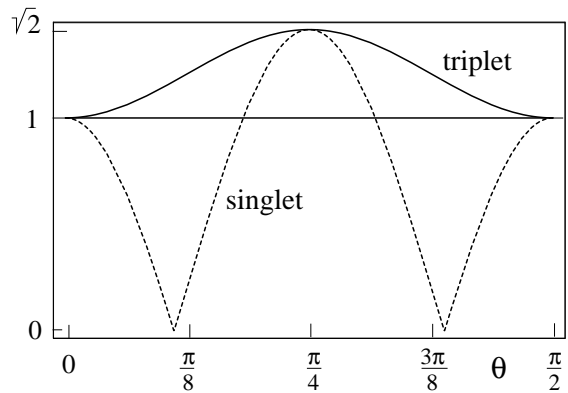}
   \caption[]{Bell inequality violation for maximal visibility $V=1$
   versus mixing angle $\theta$ (solid line: spin-triplet states;
   dashed line: spin-singlet state).}
   \label{fig:violation}
\end{figure}

We first concentrate on {\it spinless particles}; this situation can be
realized by preparing the two electrons in equal spin-states, $\chi
(\sigma_1,\sigma_2) = \delta_{\sigma_1\uparrow}\delta_{\sigma_2\uparrow}$ with
corresponding overlap $Q=1$. To begin with, we choose a plane wave form
for the wave packets with different momenta $k_1$ and $k_2$ close to $k_0$
(in order to allow for the linearized spectrum), $f(x) = \exp(ik_1 x)$ and
$g(x) = \exp(ik_2 x)$. The correlation function $E_{AB}$ takes the form
\begin{equation}
   E_{AB} = -\cos^2(2\theta) - V \sin^2(2\theta) \cos\varphi_{AB},
\end{equation}
where $\varphi_{AB} = \delta \omega\,(t_A - t_B)$ is the relative phase shift
accumulated by the two waves between the two measurement intervals and $\delta
\omega = v_0 (k_1 - k_2)$ is the frequency mismatch between the two plane
waves. The phase shift $\varphi_{AB}$ replaces the angle between the two
polarizers in the conventional Bell setup. The visibility factor $0\leq V\leq
1$ accounts for the width $\delta t$ of the time interval,
\begin{equation}
   V = \frac{\sin^2(\delta\omega \delta t/2)}{(\delta\omega \delta t/2)^2}.
\end{equation}
The other correlation functions involving intervals $A'$ and $B'$ are
obtained in the same way; their combination into the Bell inequality Eq.\
(\ref{BI}) produces a maximal violation for the angles $\varphi_{AB} =
\varphi_{A'B} = \varphi_{A'B'} = \pi/4$ and $\varphi_{AB'} =
3\pi/4$, corresponding to measurement intervals with relative distance $t_B =
t_A +\tau/8$, $t_{A'} = t_A + \tau/4$, and $\tau_{B'} = t_A +3\tau/8$,
where $\tau = 2\pi/\delta\omega$ and $t_A$ is an arbitrary reference time. In
this case the Bell inequality Eq.\ (\ref{BI}) reduces to
\begin{equation}
   {\cal E} = \bigl| V\sqrt{2} \sin^2(2\theta) + \cos^2(2\theta) \bigr|\leq 1.
   \label{BI2}
\end{equation}
For given $V$, the maximal violation ${\cal E}_{\rm max}
=V\sqrt{2}$ is reached for a symmetric beam splitter with $\theta
= \pi/4$, cf.\ Fig.\ \ref{fig:violation}. Furthermore, the
maximally allowed degree of violation ${\cal E}=\sqrt{2}$ can be
attained only for $V\approx 1$, corresponding to a short time
measurement of the current value with $\delta t < 1/v_0\delta k =
\hbar/\delta\varepsilon$: hence, the maximal violation of the Bell
inequality can be obtained for indistinguishable particles, while
a time interval with length beyond Heisenberg's uncertainty bound
$\delta t > \hbar/\delta\varepsilon$ allows for a distinction
between the two particles and the Bell inequality cannot be
violated in this classical situation. For $V \leq 1/\sqrt{2}$ the
Bell inequality is always satisfied. Within the region
$1/\sqrt{2}<V\leq 1$, the Bell inequality~(\ref{BI2}) is always
violated for any mixing angle $0<\theta<\pi/2$, although to a
lesser degree then in the symmetric point $\theta = \pi/4$.

Let us discuss the physical origin of the violation. The correlator Eq.\
(\ref{corr}) measured in the Bell test is finite, provided that both electrons
are detected within the time windows $A$ and $B$; in this case, it is
proportional to the probability $P_{\alpha_1\alpha_2}(AB)$. Although,
formally, the electrons have different energies $\varepsilon$ and thus are
distinguishable in principle, given a small time resolution $\delta t <
\hbar/\delta\varepsilon$ of the local current measurements one cannot
distinguish between the energies $\varepsilon_1=\hbar v_0 k_1$ and
$\varepsilon_2=\hbar v_0 k_2$. Under this circumstances the electrons indeed
can be considered as indistinguishable particles. Then, according to the rules
of quantum mechanics, there are two quantum alternatives contributing to a
coincident detection of the electrons in $A$ and $B$: either the electrons
with energies $\varepsilon_1$  and $\varepsilon_2$ are detected in the time
windows $A$ and $B$, respectively, or vice a versa. These two alternatives
contribute to the measurement outcome with different phases: in the first
case, the phase factor acquired by the two-particle wave function after the
first measurement at $t_A$ due to the propagation of the second particle until
$t_B$ is given by $\exp[-i\varepsilon_2(t_B-t_A)]$, while in the second case
this phase assumes the value $\exp[-i\varepsilon_1(t_B-t_A)]$. The phase
difference between the two alternatives leads to quantum interference and a
corresponding oscillatory dependence (with frequency $\delta\omega=
(\varepsilon_2 -\varepsilon_1)/\hbar$) of the probability
$P_{\alpha_1\alpha_2} (AB)$ as a function of time, with an amplitude
proportional to the visibility factor $V$. The precise bound on $\delta t$
allowing for a violation of the Bell inequality is given by $1/\sqrt{2}<V\leq
1$ or $(\varepsilon_2-\varepsilon_1)\delta t \leq 2\hbar$, corresponding to a
measurement where the Heisenberg uncertainty principle for energy--time
variables is violated.

\subsection{Space-separated domains}

\begin{figure}[bt]
   \includegraphics[width=7.5cm]{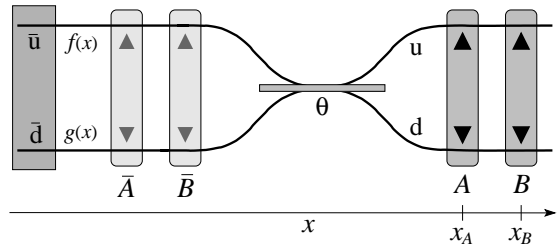}
   \caption[]{Particles incident in leads $\bar\mathrm{u}$ and $\bar\mathrm{d}$
   with wave functions $f(x)$ and $g(x)$ are mixed in a four-terminal splitter
   and analyzed through measurement of equal-time correlations within the space
   intervals $A$ and $B$ centered around $x_A$ and $x_B$. The interchangeability
   of projection (to the intervals A and B) and mixing allows to shift the
   measurement intervals to the positions $\bar{A}$ and $\bar{B}$ in the
   incoming leads; provided that the measurements in $\bar{A}$ and $\bar{B}$
   and in $A$ and $B$ are ballistically delayed in time, the measurement outcome
   is the same.}
   \label{fig:setup_b}
\end{figure}

In order to understand better the nature of the entanglement
observed in (\ref{BI2}), we consider a slightly different
experiment, where instead of using time-separated detection
intervals, the two observers Alice and Bob are measuring the
simultaneous appearance of particles in spatially separated
regions $A$ and $B$ of the setup, see Fig.\ \ref{fig:setup_b}; for
particles with a linear dispersion, these two experiments are
equivalent since a time delayed measurement with $\delta t =
t_2-t_1$ at the point $x$ corresponds to a coincident measurement
at time $t$ with $\delta x = (t_2-t_1)v_0$.  We first concentrate
on plane-wave incoming states, where the present setup with
spatially separated detectors provides additional insights. In
particular, we will see that it is the projection of the
non-entangled incoming state onto the two domains $A$ and $B$ that
defines a bipartition of the system with respect to which the lead
index becomes entangled. On the other hand, in order to perform a
Bell inequality check, we need a set of local `rotations' of the
measurement apparatus: in our setup, the parameters generating a
suitable set of local `rotations' are determined by the distance
between the measurement domains $A$ and $B$ and by the mixing
angle $\theta$.

A central element in our discussion below is the
interchangeability of mixing $U\otimes U$ and projection ${\cal
P}_{AB}$ onto the domains $A$ and $B$, where $U$ denotes the
one-particle scattering matrix of the beam splitter and the tensor
product $U\otimes U$ acts on our two-particle state. This
interchangeability is a trivial consequence of these two
operations affecting different degrees of freedom, coordinates
$x_1$ and $x_2$ and lead indices u and d. In terms of these
operators, we can relate the incoming and outgoing states via
\begin{equation}\label{eq:interch_1}
      |\Psi^\mathrm{out}_{AB}\rangle = {\cal P}_{AB} U \otimes U
      |\Psi^\mathrm{in}\rangle.
\end{equation}
Assuming that the projections onto $A$, $B$ in the outgoing leads and onto
$\bar{A}$, $\bar{B}$ in the incoming leads (see Fig.\ \ref{fig:setup_b}) are
ballistically separated (i.e., the measurement in $A$, $B$ involves the
appropriate ballistic delay time) we can write
\begin{equation}\label{eq:interch_2}
      |\Psi^\mathrm{out}_{AB}\rangle
      = U \otimes U {\cal P}_{\bar{A}\bar{B}} |\Psi^\mathrm{in}\rangle.
\end{equation}
Hence, in our discussion we are free to interchange the two operations
of mixing and projection.

Consider then an incoming state (before mixing) with single-particle wave
functions $f(x)=e^{ik_1x}$ and $g(x)=e^{ik_2x}$ with shifted momenta.  The
state incident from leads $\bar\mathrm{u}$ and $\bar\mathrm{d}$ can be
written as a simple Slater determinant,
\begin{equation}
      \label{2wave}
      |\Psi^\mathrm{in}\rangle = \int dx_1 dx_2\, f(x_1)g(x_2)\,
      \hat\psi_{\bar\mathrm{u}}^\dagger(x_1)
      \hat\psi_{\bar\mathrm{d}}^\dagger(x_2)|0\rangle,
\end{equation}
and thus is non-entangled. The lead index $\mathrm{x} \in \{\bar\mathrm{u},
\bar\mathrm{d}\}$ of the electron field operator $\hat\psi_{\mathrm{x}}$
is conveniently regarded as a pseudo-spin.

To start with, we analyze the coincident detection of two particles within the
non-overlapping regions $\bar A$ and $\bar B$ of the incoming leads, see Fig.\
\ref{fig:setup_b}, and select only those events, where each of the observers
(Alice in $\bar A$ and Bob in $\bar B$) finds only one particle. For two
particles, this projection can be described by the operator ${\cal P}_{\bar{A}
\bar{B}} = \hat N(\bar{A}) \hat N(\bar B)$, with the  particle number operator
$\hat N(X) = \int_X dx\, ( \hat \psi_{\bar\mathrm{u}}^\dagger(x)
\hat\psi_{\bar\mathrm{u}}(x) + \hat\psi_{\bar\mathrm{d}}^\dagger(x)
\hat\psi_{\bar \mathrm{d}}(x))$ counting particles in the region $X$ of the
incoming leads.  Projecting the incoming state~(\ref{2wave}) one arrives at
the state
\begin{eqnarray}
      &&|\Psi^\mathrm{in}_{\bar{A}\bar{B}} \rangle =
      \int dx_1 dx_2\, [f_{\bar{A}}(x_1) g_{\bar{B}}(x_2)
      \label{2prj} \\
      &&\qquad\quad
      + f_{\bar{B}}(x_1) g_{\bar{A}}(x_2)]\,
      \hat\psi_{\bar\mathrm{u}}^\dagger(x_1)
      \hat\psi_{\bar\mathrm{d}}^\dagger(x_2)|0\rangle,
      \nonumber
\end{eqnarray}
where $f_X(x)$ and $g_X(x)$ are equal to $f(x)$ and $g(x)$ for $x \in X$ and
vanishing outside. This projected state is no longer a simple Slater
determinant and describes a two-particle state entangled in the lead indices
and shared between the regions $\bar{A}$ and $\bar{B}$ of the incoming leads.
It is instructive to rewrite the state~(\ref{2prj}) in a pseudo-spin notation:
Assuming for simplicity that the intervals $\bar A$ and $\bar B$ are reduced
to individual points $x_{\bar A}$ and $x_{\bar B}$ we have
\begin{equation}
      |\Psi^\mathrm{in}_{\bar{A}\bar{B}}\rangle
      \propto e^{i\varphi_{\bar{A}\bar{B}}/2}
      |\!\uparrow\rangle_{\bar A} |\!\downarrow\rangle_{\bar B}+
      e^{-i\varphi_{\bar{A}\bar{B}}/2} |\!\downarrow\rangle_{\bar A}
      |\!\uparrow\rangle_{\bar B},
      \label{2prj_s}
\end{equation}
where $|\!\uparrow\rangle_X$ and $|\!\downarrow\rangle_X$ denote states of
particles localized in $X$ and residing in lead $\bar\mathrm{u}$ and
$\bar\mathrm{d}$, respectively; the orbital part of the wave function
contributes the phase factors $\exp(\pm i\varphi_{\bar{A}\bar{B}}/2)$ with
$\varphi_{\bar{A}\bar{B}} = \delta k (x_{\bar A}- x_{\bar B})$, where $\delta
k = k_1-k_2$ is a momentum mismatch.  The projected state~(\ref{2prj_s}) is in
fact maximally entangled in the lead- or pseudo-spin index with respect to
bipartitioning the system between the regions $\bar A$ and $\bar B$. In the
following, we wish to detect this entanglement in a Bell test.

The implementation of a Bell test relies on the ability to locally change the
pseudo-spin basis of the particles. To do so, we transmit the original
incoming state Eq.\ (\ref{2wave}) through a beam splitter before measuring the
presence of particles in the intervals $A$ and $B$, now located in the leads u
and d to the right of the mixer, see Fig.\ \ref{fig:setup_b}; the mixing then
acts as an equal rotation of the (pseudo-spin) basis $\bar\mathrm{u},
\bar\mathrm{d}$ for both particles. However, such a global rotation of the
original basis alone is not sufficient to perform the Bell test, as {\it
locally distinct} rotations are required as well; the latter are implemented
through different choices in the separation $\delta x = x_2-x_1$ between the
regions $A$ and $B$. Exploiting the interchangeability of projection and
mixing, cf.\ Eqs.\ (\ref{eq:interch_1}) and (\ref{eq:interch_2}), we see that
this change in distance results in a relative rotation with the angle
$\varphi_{\bar{A}\bar{B}}=\varphi_{AB}$ around the original $(\bar\mathrm{u},
\bar\mathrm{d})$-polarization axis of the pseudo spins, see
Eq.~(\ref{2prj_s}).  Writing the outgoing state (\ref{eq:interch_2}) in
pseudo-spin notation, we obtain the expression
\begin{eqnarray}
      &&|\Psi^\mathrm{out}_{AB}\rangle = -
      \cos(\varphi_{AB}/2)\sin(2\theta) \frac{|\!\uparrow\rangle_A
      |\!\uparrow\rangle_B - |\!\downarrow\rangle_A
      |\!\downarrow\rangle_B}{\sqrt{2}}
      \nonumber
      \\
      &&\quad+
      \cos(\varphi_{AB}/2) \cos(2\theta) \frac{|\!\uparrow\rangle_A
      |\!\downarrow\rangle_B + |\!\downarrow\rangle_A
      |\!\uparrow\rangle_B}{\sqrt{2}}
      \nonumber\\
      &&\quad+i\sin(\varphi_{AB}/2) \frac{|\!\uparrow\rangle_A
      |\!\downarrow\rangle_B - |\!\downarrow\rangle_A
      |\!\uparrow\rangle_B}{\sqrt{2}}.
\end{eqnarray}
This projected state describes two spatially separated localized particles
with entangled pseudo spin indices. Choosing different space separations
between the regions $A$ and $B$ allows one to change the phase $\varphi_{AB}$
and mixing by $U\otimes U$ generates a second rotation parametrized by the
angle $\theta$. Calculating the joint probabilities $P_{\alpha_1\alpha_2}(AB)
\propto |\langle \alpha_1\alpha_2| \Psi^\mathrm{out}_{AB} \rangle|^2$ for the
four settings $\alpha_1\alpha_2 \in \{\mathrm{u}\mathrm{u},
\mathrm{u}\mathrm{d}, \mathrm{d}\mathrm{u}, \mathrm{d}\mathrm{d}\}$ we find
the Bell correlation functions $E_{AB}$ as given by (\ref{Bcorr2}) and
choosing appropriate angles $\varphi_{AB}, \varphi_{AB'}, \varphi_{A'B},
\varphi_{A'B'}$ and $\theta$ one finds the Bell inequalities violated.

\subsection{Density matrix formulation}

In a last step, we reformulate our analysis in terms of density
matrices and express the Bell inequality in terms of concurrences
of density matrices reduced after projection to the intervals $A$
and $B$.  We rewrite the projected state~(\ref{2prj}) incident
from leads $\bar\mathrm{u}$ and $\bar\mathrm{d}$ in pseudo spin
representation, $|\Psi_{\bar{A}\bar{B}}\rangle = \int dx_1 dx_2\,
|\Psi_{\bar{A}\bar{B}} (x_1,x_2)\rangle$, where
\begin{eqnarray}
      &&|\Psi_{\bar{A}\bar{B}}(x_1,x_2)\rangle =
      f_{\bar A}(x_1)g_{\bar B}(x_2)\,
      |\!\uparrow\rangle_{\bar A} |\!\downarrow\rangle_{\bar B}
      \nonumber\\
      &&\qquad\qquad + f_{\bar B}(x_1) g_{\bar A}(x_2)\,
      |\!\downarrow\rangle_{\bar A} |\!\uparrow\rangle_{\bar B}.
      \label{2wave_1}
\end{eqnarray}
The joint measurement of the pseudo-spin index in the regions $\bar A$ and
$\bar B$ is described by the coordinate-reduced two-particle density operator
$$
   \bar\rho_{\bar{A}\bar{B}} \propto
   \int dx_1 dx_2\, |\Psi_{\bar{A}\bar{B}}(x_1,x_2)\rangle
   \langle \Psi_{\bar{A}\bar{B}}(x_1,x_2)|.
$$
We introduce the two-particle pseudo-spin basis
$\{|\!\!\uparrow\uparrow\rangle,$ $|\!\!\uparrow\downarrow\rangle,
|\!\!\downarrow\uparrow\rangle, |\!\!\downarrow\downarrow\rangle\}$, where
the first (second) arrow refers to the particle localized in $\bar
A$ ($\bar B$); the normalized density matrix then assumes the form
\begin{equation}
      \bar \rho_{\bar{A}\bar{B}} \!=\! \frac{1}{F_{\bar{A}}G_{\bar{B}}\!
                                             +\!G_{\bar{A}}F_{\bar{B}}}
      \left( \begin{array}{cccc}
      0&0&0&0\\
      0& F_{\bar{A}}G_{\bar{B}} & - S_{\bar{A}} S_{\bar{B}}^*& 0\\
      0& -S_{\bar{A}}^* S_{\bar{B}}& G_{\bar{A}}F_{\bar{B}}& 0\\
      0&0&0&0
      \end{array} \right)\!,
      \label{rAB}
\end{equation}
where $F_X=\int_X |f(x)|^2 dx$, $G_X = \int_X |g(x)|^2 dx$ and $S_X = \int_X
f(x)g^*(x) dx$ with $X\in\{\bar{A},\bar{B}\}$.  Although initially the two
particles have been in a pure state, the {\it reduced} density matrix
(\ref{rAB}) corresponds to a mixed state with $\bar{\rho}_{\bar{A}
\bar{B}}^2\neq \bar \rho_{\bar{A}\bar{B}}$. The calculation of the
entanglement in the mixed two-particle state Eq.\ (\ref{rAB}) corresponds to
finding the concurrence ${\cal C}(\bar \rho_{\bar{A}\bar{B}})$ of a two-qubit
problem, and thus can be calculated following the scheme introduced by
Wootters~\cite{wootters}, ${\cal C}(\bar \rho_{\bar{A}\bar{B}})=\max\{ 0,
\sqrt{\lambda}_1 - \sqrt{\lambda_2} -\sqrt{\lambda_3} - \sqrt{\lambda_4}\}$
where $\lambda_1\geq \lambda_2\geq \lambda_3 \geq \lambda_4\geq 0$ are the
eigenvalues of the matrix $\bar\rho_{\bar{A}\bar{B}} \bar q_{\bar{A}\bar{B}}$
with $\bar q_{\bar{A}\bar{B}} = (\sigma_y\otimes \sigma_y) \bar
\rho_{\bar{A}\bar{B}}^* (\sigma_y\otimes \sigma_y)$, $\sigma_y$ is a Pauli
matrix, and $\otimes$ denotes the tensor product. The result of this
calculation provides us with the expression
\begin{equation}\label{eq:conc}
   {\cal C}(\bar \rho_{\bar{A}\bar{B}}) =
   \frac{2|S_{\bar{A}}||S_{\bar{B}}|}
   {F_{\bar{A}}G_{\bar{B}}+G_{\bar{A}}F_{\bar{B}}}.
\end{equation}
This quantity is indeed restricted to the interval $[0,1]$, as
follows from the Cauchy-Schwartz inequality and the inequality
$(\sqrt{F_{\bar{A}} G_{\bar{B}}}-\sqrt{F_{\bar{A}} G_{\bar{B}}})^2
> 0$, $2|S_{\bar{A}}||S_{\bar{B}}| \leq 2
\sqrt{F_{\bar{A}}G_{\bar{A}} F_{\bar{B}}G_{\bar{B}}} \leq
F_{\bar{A}} G_{\bar{B}}+G_{\bar{A}}F_{\bar{B}}$. The state
described by Eq.\ (\ref{rAB}) is trivial (i.e., not entangled or
classical) only for zero overlap $S_{\bar{A}}=0$ and/or
$S_{\bar{B}}=0$. Physically, the vanishing of the overlap between
the wave functions $f(x)$ and $g(x)$ in either of the two regions
$\bar{A}$ and $\bar{B}$ implies, that these orbital states are
{\it perfectly distinguishable} via a local measurement. In this
situation the corresponding density matrix $\bar
\rho_{\bar{A}\bar{B}}$ can be written in a convex form $\bar
\rho_{\bar{A}\bar{B}} = \sum_i p_i \bar\rho_{\bar{A}}^{(i)}
\otimes \bar \rho_{\bar{B}}^{(i)}$, with probabilities $p_i\geq 0$
and $\sum_i p_i=1$, and thus is separable. On the other hand, for
$S_{\bar{A},\bar{B}} \neq 0$ one cannot perfectly distinguish
between the different orbital states via local measurements,
resulting in an interference between the different terms of the
anti-symmetric wave function Eq.\ (\ref{2wave_1}), a non-separable
density matrix, and a finite concurrence.

Next, we analyze the reduced density matrix in the outgoing leads, i.e., in
the new basis $\{\mathrm{uu}, \mathrm{ud}, \mathrm{du}, \mathrm{dd}\}$.
Exploiting the possibility of exchanging real space projection and mixing,
we can simply rotate the projected density matrix according to
\begin{equation}\label{rho_AB}
      \rho_{AB}=(U\otimes U)\bar\rho_{\bar{A}\bar{B}}(U^\dagger \otimes
      U^\dagger).
\end{equation}
The diagonal elements of the density matrix $\rho_{AB}$ directly provide the
detection probabilities $P_{\alpha_1\alpha_2}(AB)$, which then can be used in the
calculation of the Bell correlation function $E_{AB}$, Eq.~(\ref{Bcorr}),
\begin{equation}\label{eq:EABC}
   E_{AB}=-\cos^2(2\theta)-\sin^2(2\theta)\,{\cal C}(\rho_{AB})\cos\varphi_{AB},
\end{equation}
where the angle $\varphi_{AB}$ is given by the overlap integrals, $\varphi_{AB}
= \arg (S_A S_B^*)$; combining Eqs.\ (\ref{eq:conc}) and (\ref{eq:EABC}) we
immediately recover the original expression (\ref{Bcorr2}). Choosing four
different intervals $A$, $B$, $A'$, and $B'$ (note that the selection of these
intervals is non-trivial in the general situation discussed here, as
$\varphi_{AB}$ now involves overlap integrals), we can set up the Bell
inequality~(\ref{BI}) and find the result expressed in terms of concurrences
${\cal C}_{A'B} = {\cal C}(\rho_{A'B})$,
\begin{eqnarray}
   &&\bigl| \sin^2(2\theta)
   \bigl[ {\cal C}_{AB} \cos\varphi_{AB}
   - {\cal C}_{AB'}\cos\varphi_{AB'}
   \\ &&\>\>
   + {\cal C}_{A'B} \cos\varphi_{A'B}
   + {\cal C}_{A'B'}\cos\varphi_{A'B'}\bigr]
   + 2\cos^2(2\theta) \bigr|\leq 2.
   \nonumber
\end{eqnarray}
Choosing plane waves for $f(x)$ and $g(x)$, the concurrences take the value
${\cal C}_{AB} = {\cal C}_{AB'} = {\cal C}_{A'B}= {\cal C}_{A'B'} = V$ and the
Bell inequality reduces to the simpler form found earlier, see Eq.\
(\ref{BI2}); for the general case, the degree of violation depends separately
on the shapes $f(x)$ and $g(x)$ of the orbital wave functions in each region
$A$, $A'$, $B$, and $B'$ via the corresponding concurrences.

\subsection{Particles with spin}

So far, we have considered only spinless particles or, more exactly, two
electrons in a spin-triplet state with the same spin polarization of the
electrons, $\chi^\mathrm{tr}_{+1}(\sigma_1,\sigma_2) =
\delta_{\sigma_1\uparrow} \delta_{\sigma_2\uparrow}$ and
$\chi^\mathrm{tr}_{-1}(\sigma_1,\sigma_2) = \delta_{\sigma_1\downarrow}
\delta_{\sigma_2\downarrow}$. Since all spin-dependence of the Bell inequality
is encoded in the overlap $Q$ of the spin wave-functions, see Eq.\
(\ref{Bcorr2}), one concludes that the above results are valid as well for the
third maximally entangled triplet state, $\chi^\mathrm{tr}_0
(\sigma_1,\sigma_2) =(\delta_{\sigma_1\uparrow} \delta_{\sigma_2\downarrow} +
\delta_{\sigma_1 \downarrow} \delta_{\sigma_2 \uparrow} )/\sqrt{2}$ with
$Q=1$. On the other hand, the character of violation is modified for the
spin-singlet state $\chi^\mathrm{sg}(\sigma_1, \sigma_2) = (\delta_{\sigma_1
\uparrow} \delta_{\sigma_2\downarrow} - \delta_{\sigma_1 \downarrow}
\delta_{\sigma_2 \uparrow} )/\sqrt{2}$ with $Q=-1$. Choosing a set of optimal
time intervals $A$, $B$, $A'$, and $B'$, the resulting Bell inequality takes
the form
\begin{equation}
   {\cal E} = \bigl| V\sqrt{2}\sin^2(2\theta) -\cos^2(2\theta)\bigr|\leq 1.
\end{equation}
The main difference to the previous result for spin-triplet states is that
this inequality can be violated only for a sufficiently large visibility
factor $1/\sqrt{2}<V\leq 1$ and a beam splitter with a mixing angle $\theta$
sufficiently close to optimal, $\theta \in[\pi/4 - \theta_c, \pi/4+\theta_c]$,
where the critical angle $\theta_c$ is given by $\sin^2(2\theta_c) =
1/(1+V\sqrt{2})$, cf.\ Fig.\ \ref{fig:violation}.  This result allows one to
distinguish between triplet and singlet incoming states by measuring a Bell
inequality involving only orbital degrees of freedom, see also Ref.\
\onlinecite{loss_00}.

Moreover, assuming that the incident electrons have opposite spin
polarization, i.e., their spin state can be written as a superposition
$\chi^\mathrm{in} = \alpha \chi^\mathrm{tr}_0 + \beta\chi^\mathrm{sg}$, the
degree of violation of the orbital Bell inequality gives a lower bound on the
value of the concurrence in the spin part of the wave function. Indeed, in
this case $Q(\chi^\mathrm{in}) = |\alpha|^2 - |\beta|^2$ and the maximal
violation of the orbital Bell inequality for $V=1$ and symmetric scattering is
given by ${\cal E}_\mathrm{max} = |Q|\sqrt{2}$. At the same time, the
concurrence ${\cal C}(\chi^\mathrm{in}) = |\alpha^2-\beta^2| \geq |Q|$, with
equality established for real $\alpha$ and $\beta$ (note that ${\cal
C}(\chi^\mathrm{in})$ gives the degree of (useful) spin entanglement in the
outgoing leads u and d). Hence, measuring the entanglement ${\cal
E}_\mathrm{max}$ of the orbital part of the wave function (which leaves the
spin component untouched), provides a (lower) estimate of the degree of spin
entanglement of the incoming state.  If the spin wave function of the incoming
electrons factorizes, $\chi^\mathrm{in} (\sigma_1,\sigma_2) =
\delta_{\sigma_1\uparrow} \delta_{\sigma_2\downarrow}$, the orbital Bell
inequality never can be violated since in this situation the electrons are
distinguishable and thus the detection of an electron with given spin in one
of the outgoing leads always allows to determine its origin.
\begin{figure}[bt]
   \includegraphics[width=7.5cm]{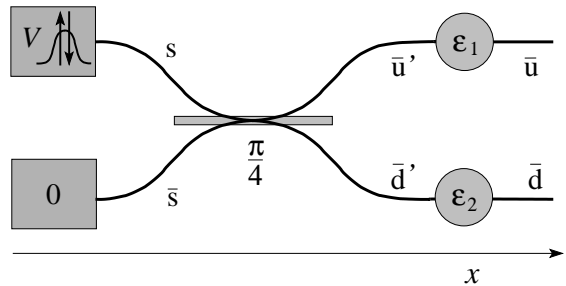} \caption[]{Spin-singlet source: The
   voltage pulse $V(t)$ injects a singlet-pair of electrons into the lead s;
   the $\pi/4$ four-terminal splitter distributes the particles with equal
   probabilities among the two leads $\bar\mathrm{u}'$ and $\bar\mathrm{d}'$.
   The resonances in the quantum dots select the desired energies
   $\varepsilon_1$ and $\varepsilon_2$; dots residing in the Coulomb
   blockade regime inhibit the propagation of two electrons into the same lead, such
   that the two-particle incident state involves one particle in each of the leads
   $\bar\mathrm{u}$ and $\bar \mathrm{d}$.} \label{fig:source}
\end{figure}

\section{Spin-singlet/triplet sources}

Finally, we discuss the potential experimental realization of the proposed Bell
test. The source of spin-entangled incoming particles can be realized with the
help of a beam splitter, followed by leads with dots serving as energy filters
defined through resonance levels at energies $\varepsilon_1$ and
$\varepsilon_2$, see Fig.\ \ref{fig:source}. We assume that both dots reside
in the strong Coulomb blockade regime; applying a single-electron voltage
pulse to the source lead $\mathrm{s}$, two electrons in a singlet state are
detached from the Fermi see~\cite{lesovik,levitov}. There is only one
scattering process, involving trajectories where the electrons tunnel through
different quantum dots, for which the two electrons reach the second beam
splitter. The incoming state in the leads $\bar\mathrm{u}$ and
$\bar\mathrm{d}$ then is of a spin-singlet type with different energies
$\varepsilon_1$ and $\varepsilon_2$ as defined through the dot resonances.
All other scattering processes with only one or no electrons propagating towards
the second beam splitter are irrelevant as they do not contribute to the
correlation measurement.

A spin entangled triplet state can be generated with the help of
spin-polarized reservoirs with polarizations $\uparrow$ and $\downarrow$
attached to the leads $\mathrm{s}$ and $\bar\mathrm{s}$, respectively.
Applying a single-electron voltage-pulse to each reservoir, two electrons with
opposite spins are injected into the leads $\mathrm{s}$ and $\bar \mathrm{s}$,
see Fig.\ \ref{fig:source}. The state $|\Psi_{\rm s\bar s}\rangle = \int dx_1
dx_2\, f(x_1) g(x_2) \hat\psi^\dagger_{\mathrm{s}\uparrow }(x_1) \hat
\psi^\dagger_{\bar\mathrm{s} \downarrow}(x_2)|0\rangle$ incident on the
symmetric beam splitter emerges with a component
\begin{eqnarray}
      |\Psi_{\rm \bar u' \bar d'}\rangle \!\!&\propto&\!\!\! \int \! dx_1 dx_2
      \, \bigl[ g(x_1) f(x_2)\, \hat\psi_{\bar \mathrm{u}'
      \downarrow}^\dagger(x_1)
      \hat \psi_{\bar\mathrm{d}' \uparrow}^\dagger(x_2)
      \label{Psi_ud}
      \\
      &&\qquad +f(x_1)g(x_2)\,
      \hat \psi_{\bar\mathrm{u}' \uparrow}^\dagger(x_1)
      \hat \psi_{\bar\mathrm{d}' \downarrow}^\dagger(x_2)
      \bigr] |0\rangle\nonumber
\end{eqnarray}
describing electrons scattered into different leads $\bar\mathrm{u}'$ and $\bar
\mathrm{d}'$; it is this component which can propagate through the subsequent
energy filter and contribute to the current correlators. The propagation of this
component through the quantum dots results in an entangled spin-triplet state
of the form given by Eq.\ (\ref{inst}) with $f(x) = \exp(ik_1 x_1)$ and
$g(x)=\exp(ik_2 x)$.

Above, we have considered an idealized situation where only two electrons are
present in the system, while in a realistic situation one deals with
electronic reservoirs at finite temperature.  The associated equilibrium
fluctuations then generate noise signals which are of the same order as the
correlations associated with the injection of the two electrons. We note,
however, that the corresponding equilibrium current correlators $\langle \hat
I_{\alpha_1}(x,t_1) \hat I_{\alpha_2}(x,t_2) \rangle_\mathrm{eq}$ assume significant
values only for {\it instantaneous} or {\it ballistically retarded} variables,
i.e., at times $t_2=t_1$ and $t_2-t_1=2\ell/v_{\rm\scriptscriptstyle F}$ in
the same leads and $t_2-t_1=2\ell/v_\mathrm{F}$ in opposite leads (here,
$\ell$ denotes the distance between the position of measurement and the
reflecting dots). The Bell test involves correlations at time differences of
the order of $\tau=2\pi/\delta\omega$ and a proper choice of the frequency
mismatch $\delta\omega$ always allows one to render the contribution from
equilibrium fluctuations negligible. Another restriction on $\tau$ is due to
dephasing and electron-electron interactions; we then have to assume that the
characteristic times associated with these processes are larger then $\tau$.

\section{Conclusion}

We have discussed how to make use of quantum indistinguishability as a
resource to generate non-classical correlations: the indistinguishability of
particles enforces proper symmetrization of their wave function and results in
non-factorizable states. We have demonstrated how to generate such states with
the help of quantum dots residing in the Coulomb blockade regime and have
determined their degree of entanglement as measured in a Bell inequality test
based on auto- and cross-current correlators. In a real experiment, the latter
are measured over a finite time or space domain. As a result, we obtain an
interesting interplay between the measurement accuracy (time or space resolution)
and the degree of non-locality as measured in the Bell inequality test: the more
information is gained that locally distinguishes between the particles, the
smaller is the degree of violation. Once the uncertainty principle allows for
the identification of the particle, the Bell inequality cannot be violated
any longer. This feature can be exploited in the design of experiments testing
the above predictions: choosing a small energy difference $\delta\varepsilon =
\varepsilon_2-\varepsilon_1$ allows for a slow measurement with a less
stringent time resolution, while the violation of the Bell test remains
observable.  On the other hand, the energy difference $\delta \varepsilon$ has
to be chosen sufficiently large in order to avoid the influence of decoherence
or interactions.

The above setup for spinless particles provides an alternative for the
observation of the two-particle interference as proposed by Samuelsson {\it et
al.} \cite{SamBut_03} and recently observed by Neder {\it et al.}
\cite{Heiblum_07}; here, the role of the magnetic flux $\Phi$ penetrating the
Hanbury-Brown Twiss interferometer is replaced by the time-delay of subsequent
measurements in the correlator. Adding the spin degree of freedom, we are
confronted with two distict situations: if the spin degree of freedom allows
to distinguish between the particles (this is the case for the spin-state
$\chi(\sigma_1,\sigma_2) = \delta_{\sigma_1\uparrow} \delta_{\sigma_2
\downarrow}$) the Bell inequality is never violated. On the other hand,
entangled spin states in the singlet or triplet sector (these are the states
$\chi^\mathrm{sg}$ and $\chi^\mathrm{tr}_0$) can generate maximal violation of
the Bell inequality; finally, the superposition of these states reduces the
spin-entanglement and the degree of violation in the orbital Bell inequality
gives a lower bound on the spin-concurrence, with an ideal measurement
providing the best bound.

We thank Gordey Lesovik for discussions and acknowledge the financial support
from the Swiss National Foundation, the RFBR grant No.~06-02-17086-a and
the Programm ``Quantum Macrophysics'' of RAS.


\begin{thebibliography}{99}

\bibitem{review} C.\ W.\ J.\ Beenakker,
   in {\it Quantum Computers, Algorithms and Chaos}, Proceedings of the
   Int.\ School of Physics ``Enrico Fermi",
   Varenna 2005, Vol.\ {\bf 162} (IOS Press, Amsterdam, 2006);
   arXiv:cond-mat/0508488.

\bibitem{cht_00} N.\ M.\ Chtchelkatchev, G.\ Blatter, G.\ B.\ Lesovik,
   and Th.\ Martin,
   Phys.\ Rev.\ B {\bf 66}, 161320(R) (2002).

\bibitem{loss_03} G.\ Burkard, and D.\ Loss,
   Phys.\ Rev.\ Lett.\ {\bf 91}, 087903 (2003).

\bibitem{buttiker} P.\ Samuelsson and M.\ B\"uttiker,
   Phys.\ Rev.\ B {\bf 73}, 041305 (2006).

\bibitem{fazio} V.\ Giovannetti, D.\ Frustaglia, F.\ Taddei, and R.\ Fazio,
   Phys.\ Rev.\ B {\bf 74}, 115315 (2006);
   V.\ Giovannetti, D.\ Frustaglia, F.\ Taddei, and R.\ Fazio,
   Phys.\ Rev.\ B {\bf 75}, 241305(R) (2007).

\bibitem{SamBut_03} P.\ Samuelsson, E.\ V.\ Sukhorukov, and M.\ B\"uttiker,
   Phys.\ Rev.\ Lett.\ {\bf 92}, 026805 (2004).

\bibitem{Heiblum_07} I.\ Neder, M.\ Heiblum, D.\ Mahalu, and V.\ Umansky,
   Phys.\ Rev.\ Lett.\ {\bf 98}, 036803 (2007) and I.\ Neder, N.\ Ofek, Y.\
   Chung, M.\ Heiblum, D.\ Mahalu, and V.\ Umansky, Nature {\bf 448}, 333
   (2007).

\bibitem{loss_00} G.\ Burkard, D.\ Loss, and E.\ V.\ Sukhorukov,
   Phys.\ Rev.\ B {\bf 61}, R16303 (2000).

\bibitem{Lebedev_05} A.V.\ Lebedev, G.B.\ Lesovik, and G.\ Blatter,
   Phys.\ Rev.\ B {\bf 71}, 045306/1-9 (2005); note that with only
   two incident electrons there is no restriction on the accumulation
   time, see Sec. III B.

\bibitem{clauser} J.F.\ Clauser and M.A.\ Horne,
   Phys.\ Rev.\ D {\bf 10}, 526 (1974).

\bibitem{aspect} A.\ Aspect, P.\ Grangier, and G.\ Roger,
   Phys.\ Rev.\ Lett.\ {\bf 49}, 91 (1982).

\bibitem{transport} G.\ B.\ Lesovik, JETP Lett.\ {\bf 49}, 592 (1989);
   M.\ B\"uttiker, Phys.\ Rev.\ Lett.\ {\bf 65} 2901 (1990);
   Ya.\ M.\ Blanter and M.\ B\"uttiker,  Phys.\ Rep.\ {\bf 336}, 1 (2000).

\bibitem{wootters} W.\ K.\ Wootters,
   Phys.\ Rev.\ Lett.\ {\bf 80}, 2245 (1998).

\bibitem{lesovik} L.\ S.\ Levitov, H.\ W.\ Lee, and G.\ B.\ Lesovik,
   J.\ Math.\ Phys.\ {\bf 37}, 4845 (1996).

\bibitem{levitov} J.\ Keeling, I.\ Klich, and L.\ S.\ Levitov,
   Phys.\ Rev.\ Lett.\ {\bf 97}, 116403 (2006).

\end{thebibliography}
\end{document}